\documentclass[aps,nofootinbib]{revtex4}
\usepackage{graphicx}

\begin{document}

\title{Driven diffusive systems of active filament bundles}

\author{P. K. Mohanty$^{1,2,}$}
\email{pk.mohanty@saha.ac.in}  
\author{K. Kruse$^{1,}$}
\email{karsten@pks.mpg.de}
\affiliation{
              $^1$Max-Planck-Institute for the Physics of Complex Systems,
              N\"othnitzer Str. 38,
              01187 Dresden,
              Germany\\
             $^2$Theoretical Condensed Matter Physics Division,
 Saha Institute of Nuclear Physics,
 1/AF Bidhan Nagar, Kolkata-700 064, India}

\begin{abstract}
The cytoskeleton is an important subsystem of cells that is involved 
for example in cell division and locomotion. It consists of filaments
that are cross-linked by molecular motors that can induce relative
sliding between filaments and generate stresses in the network.
In order to study the effects of fluctuations on the dynamics of such a 
system we
introduce here a new class of driven diffusive systems mimicking 
the dynamics of active filament bundles where the filaments are
aligned with respect to a common axis. After introducing 
the model class we first analyze an exactly solvable case and find
condensation. For the general case we perform a mean-field analysis
and study the behavior on large length scales by coarse-graining.
We determine conditions for condensation and establish a relation
between the hopping rates and the tension generated in the bundle.
\end{abstract}
\maketitle
\section{Introduction}
\label{sec:intro}

Our current understanding of non-equilibrium systems owes a lot
to the studies of driven diffusive systems~\cite{schm95}. These are models of 
particles hopping from site to site on a lattice where the hopping rates
do not fulfill the condition of detailed balance. Even though the
rules governing the stochastic dynamics  are often 
very simple, these models show a variety of phenomena that are 
unknown to systems at thermodynamic equilibrium. In particular, 
even in one spatial dimension they can exhibit phase 
separation and phase transitions. 

Biology provides ample examples of systems that are maintained
out of thermodynamic equilibrium and physical studies of these 
systems are essential for understanding their behavior. In contrast 
to the driven diffusive systems mentioned before, they in general
do not follow simple rules. This can be illustrated by the cytoskeleton
which is a network of filamentous proteins, mainly actin-filaments
and microtubules~\cite{albe02,bray01,howa01}. 
The cytoskeleton plays an essential role in many vital cellular 
processes as, for example, cell division and locomotion.
Cytoskeletal filaments are polar objects, because they have two 
structurally distinguishable ends. Other proteins associated with
the cytoskeleton control filament assembly and disassembly.
Furthermore, they can form cross-links between filaments. In addition
to passive cross-links there are active cross-links formed by
motor proteins. These are molecules that can move 
directionally along filaments. Their motion is driven by the
hydrolysis of Adenosine-Tri-Phosphate during which a phosphate 
group is cleaved from the molecule. The chemical energy thus 
liberated is then transformed by the motor proteins into mechanical 
work. The direction of their motion is determined by the orientation
of the filament. If a motor protein connects two filaments its motion 
results in relative motion between the filaments and creates stresses 
in the filament network. 

Now each of the proteins involved in the dynamics of the cytoskeleton
is already a complicated system
on its own, consisting of many atoms. However, the separation of time
scales allows phenomenological descriptions of the slow modes. 
Essential features of cytoskeletal dynamics can be studied using
simple models. For example, driven diffusive systems have been 
used to study the growth  of filaments against external 
forces~\cite{stuk04}, the motion of single 
motors on a filament~\cite{kolo98,fish99}, as well as collective effects 
of motor molecules interacting through excluded volume 
effects~\cite{lipo01,krus02,parm03}. 

Motivated by the dynamics of bundles of polar filaments interacting
with molecular motors, we study in this work a new class of driven 
diffusive systems. In contrast to commonly studied models, elementary 
events consist here of 
a correlated motion of two particles reflecting the effects of filament
motor interactions. After defining the model class we study 
a sub-class that allows for an exact solution of the steady state.
We identify criteria for condensation in these models. 
In a filament bundle, condensation corresponds to bundle shortening.
We then analyze cases without exact solution by employing a
mean-field approximation. Finally, we look at a coarse-grained
version and relate the model to phenomenological descriptions
of active filament bundles. In this way we can connect the kinetic
hopping rates defining the driven diffusive system to the stress
generated in a filament bundle. 

\section{Active filament bundles} 
\label{sec:model}

We will now briefly review a microscopic physical model of the filament 
dynamics in a bundle induced by motor molecules. Motivated by this 
model we will then define the processes that determine the class
of driven diffusive systems to be studied below.

\subsection{The minimal model}
\label{sec:minimal}
Consider a bundle of polar filaments in which all filaments are
aligned with respect to a common axis. Each filament has one of
two possible orientations which in the following will be referred to
as ``plus'' and ``minus'', respectively. Filament pairs are set into 
relative motion by the action of molecular motors which form active 
cross-links. As a consequence, the bundle can shorten
and filaments can be separated according to their
polarity~\cite{taki91,tana04}. 

For sufficiently large filament densities, the state of the 
bundle can be represented by two densities
corresponding to the 
distributions of filaments oriented in one way and the other. 
They  are denoted $c^{+}$ and $c^{-}$, respectively. Due to mass 
conservation,
the time evolution of the densities is governed by continuity equations 
\begin{equation}
\label{eq:cont}
\partial_{t}c^{\pm}+\partial_{x} j^{\pm}=0
\end{equation}
The currents $j^{\pm}$ reflect various active 
processes in the bundle, for example the action of motors 
connecting filament pairs. If filaments are created and destroyed
or change their orientation,
then source terms have to be added on the right hand side of 
Eq.~(\ref{eq:cont}). 
In the following, these
processes will be neglected such that the numbers of plus- and 
minus-filaments are both conserved independently.  

Different approaches have been developed to derive expressions
for the filament currents. Phenomenological equations have been
derived based on systematic expansions of the currents in terms 
of the filament densities and their derivatives respecting the 
symmetries of the system~\cite{lee01,kim03,krus03a,krus04,krus05}.
The same
approach has been used for describing the hydrodynamics of
actively propelled particles like bacteria which share
a number of essential features with active filament systems~\cite{simh02,tone05}.
The virtue of these approaches is
that they are independent of many microscopic details of the
active processes, most of which are only partially known at present.
In more 
microscopic descriptions, expressions for the currents follow from 
an analysis
of the action of active cross-links formed by motor 
molecules~\cite{krus00,krus03,live03,live05,zumd05}. 

We will now discuss 
further the approach presented in Ref.~\cite{krus03} as it will 
form the basis for the driven diffusive systems discussed later.
It is based on an analysis of the momentum flux in the 
bundle. Since
the motion of filaments in a bundle typically occurs with velocities
in the order of $\mu$m/min, the dynamics is 
dominated by friction and inertial terms can be neglected.
The conservation of momentum is thus 
expressed by a balance of forces. Summing up all the forces
acting on a filament connected to other filaments by active motors
and moving against an immobile fluid leads to a general expression 
for the currents in terms of the internal forces. Within the framework
developed in Ref.~\cite{krus03} it is assumed that the dominant
contributions to the currents result from filament pair-interactions, 
which is for example valid in the case of a low motor density. In the
simplest case, the motor density is not considered as a dynamic 
variable, such that it only enters in the parameters describing the
dynamics.

\begin{figure}
 \centerline{ \includegraphics*[width=10.0 cm]{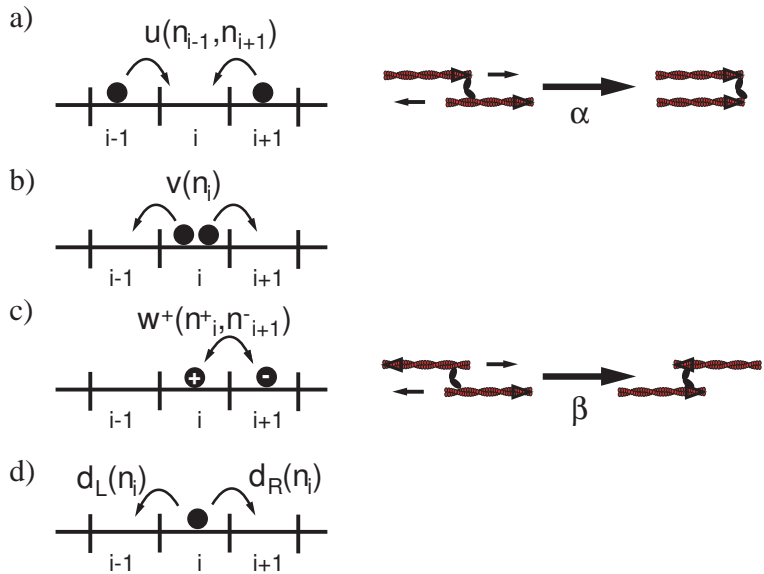}}
\caption{ The three processes governing the particle dynamics. a)
Particles of the same kind separated by one site, hop simultaneously
to the middle site with rate $u$. On the right the corresponding 
process of two filaments with the same orientation aligning their
plus-ends is illustrated; the coupling parameter in the minimal model
is $\alpha$. b) The opposite event of two particles
of the same kind leaving a site to different neighboring sites  occurs with rate $v$. c)
 Two neighboring particles of different kinds exchange their position 
 with rates $w^{\pm}$. On the right the corresponding process of 
 sliding two filaments of opposite orientation is illustrated; the
 corresponding coupling parameter in the minimal model is $\beta$.
d) Particles hop to the left or right neighboring sites 
with rates $d_{L,R}$.} 
\label{fig:process}
\end{figure}

For a simple model of the internal forces between filament pairs, the
current of plus-filaments is given by~\cite{krus00,krus03}
\begin{eqnarray}
\label{eq:jmimo}
j^{+}(x) & = &-\alpha\partial_x\int_0^\ell d\xi\;
[c^+(x+\xi)-c^+(x-\xi)]c^+(x) \nonumber\\
&&+\beta \partial_x\int_{-\ell}^\ell d\xi\; c^-(x+\xi) 
c^+(x)-D\partial_{x}c^{+}(x) + v_{\rm tr}c^{+}(x)
\end{eqnarray}
This current defines the minimal model of active filament bundles.
The first term describes the contribution to the current resulting from
interactions between filaments of the same orientation, while the second
term accounts for interactions between filaments of opposite 
orientation. The corresponding coupling constants are $\alpha$ and 
$\beta$, respectively. The integrals reflect that filaments can interact
whenever they overlap, where $\ell$ is the filament length which is
taken to be the same for all filaments. The terms resulting from the 
action of active cross-links have a simple interpretation. The 
interaction of filaments of the same orientation is such that, for 
$\alpha>0$, a filament pair tends to increase its overlap, while for 
filaments of opposite orientation the distance between their plus ends 
decreases if $\beta>0$, see Fig.~\ref{fig:process}.

The following term is a diffusion current that captures effects of
fluctuations in the system. Since the fluctuations are not only thermal,
the effective diffusion constant $D$ does not satisfy an 
Einstein-relation. Finally, the convective part $v_{\rm tr}c^{+}$ of the 
current has been introduced in Ref.~\cite{zumd05}  to 
describe filament treadmilling. In this process filament subunits
are added at the plus-end and removed with the same rate at the 
minus-end. This leads to an apparent motion of the filament in
the direction of the plus-end while the length of the filament remains
constant.

The expression for the current of minus-filaments
is obtained from expression (\ref{eq:jmimo}) by interchanging
the superscripts $+$ and $-$ and by replacing $\beta$
with $-\beta$ as well as $v_{\rm tr}$ by $-v_{\rm tr}$. This 
assures that the currents 
transform correctly under space inversion which also 
transforms plus filaments into minus filaments and vice versa.
Furthermore, the currents respect the conservation of momentum.

An analysis of the dynamic equations reveals that the important
parameter for the dynamics is $\alpha$, the coupling between
filaments of the same orientation.
If it exceeds a critical value, a homogeneous distribution
of filaments becomes unstable~\cite{krus00,krus03}. Provided
that $\beta=0$, filaments then aggregate at some position. If on the
contrary $\beta\neq0$, then oscillatory solutions in the form of 
traveling waves can be generated~\cite{krus01}. The critical value
$\alpha_{c}$ is inversely proportional to the filament density 
$c_{0}$ of the homogeneous state, $\alpha_{c}\propto c_{0}^{-1}$.

We will now use these elementary processes as a motivation
for defining a class of driven diffusive systems capturing essential 
parts of the dynamics of active filament bundles.

\subsection{Stochastic dynamics of interacting particles}

Consider a one-dimensional lattice of $L$ sites. On the lattice there 
are two kinds of particles, referred to as ``plus'' and ``minus'', which 
represent the two  possible orientations of the filaments in a bundle.
 Each site $i$ with $i=1,\ldots,L$ accommodates 
$n^{+}_{i}+n^{-}_{i}\ge0$ particles, where $n^{\pm}_{i}$ are 
the occupation numbers of plus- and minus-particles. The 
dynamics of the particles is determined by processes corresponding 
to the different terms appearing in the current (\ref{eq:jmimo}), see 
Fig.~\ref{fig:process}. 

First of all,
two particles of the same kind located respectively on sites $i-1$
and $i+1$ can simultaneously hop to site $i$. 
This process corresponds to the dynamics of a pair of equally oriented
filaments in a bundle that are cross-linked by a motor. The action of
the motor tends to increase the overlap of the filaments, 
see Fig.~\ref{fig:process}a, right. The rate of such an 
event depends on the number of particles on sites $i-1$ and $i+1$ 
and is denoted by $u(n^\pm_{i-1},n^\pm_{i+1})$ with $u(n,m)=u(m,n)$.
Note, that the function $u$ is the same for both kinds of 
particles\footnote{This process is similar to the one introduced 
in Ref.~\cite{krus00} where the ``interaction range'' was larger
and $u$ was chosen to be constant.}. For the following analysis, 
it will turn out to be convenient to allow also for the opposite events to 
occur even though there is no direct analog in the minimal model: Two
particles on site $i$ can separate and move to sites $i-1$ and $i+1$, 
respectively. The corresponding rate $v$ depends on the number of
particles at $i$, i.e. $v(n_{i})$.

The next process reflects the action of a motor on a pair of
oppositely oriented filaments, illustrated in Fig.~\ref{fig:process}c, right.
A plus particle at $i$ will exchange its position with
a minus particle at $i+1$ with a rate $w^{+}$ that depends on 
$n^{+}_{i}$ and $n^{-}_{i+1}$, i.e. $w^{+}(n^{+}_{i},n^{-}_{i+1})$.
The opposite process occurs with rate $w^{-}(n^{-}_{i}, n^{+}_{i+1})$.
Motivated by the behavior of the term proportional to $\beta$ in 
Eq.~(\ref{eq:jmimo}) under the combined operation $+\leftrightarrow-$ 
and space inversion, we require $w^{\pm}(m,n)=w^{\mp}(n,m)$.

Furthermore, we allow for processes similar to diffusion: particles
can spontaneously hop to the left or right neighboring site with
rates $d^{\pm}_{L}$ and $d^{\pm}_{R}$, respectively, each of which 
can depend on the number of particles at the site. Motivated by the
behavior of the filament system under space inversion, we require 
$d^{+}_{L}=d^{-}_{R}$ and $d^{+}_{R}=d^{-}_{L}$. Note, that if
$d^{+}_{R}\neq d^{+}_{L}$ then there is a spontaneous flux of 
particles similar to the convective term $v_{\rm tr}c^{+}$ in 
Eq.~(\ref{eq:jmimo}).
Finally, the rates have to respect that a process can 
only take place if there is a particle at the donating site(s), i.e., 
$u(n,n')=0$ if either 
$n=0$ or $n'=0$, $v(0)=v(1)=0$, and $d^{\pm}_{L,R}(0)=0$.

If only one kind of particles
is present and if $u=v=0$, the model reduces to a class of 
hopping models known as Zero Range Process (ZRP) \cite{evan05}.
In these models, the rate of a particle leaving a site depends
on its  occupation number but is otherwise unconstrained.
In spite of this simplicity, phase transitions are found in this class
even in one dimension. 
The ZRP allows for an analytical solution of the steady state.
This property is intimately linked to the factorization of the steady
state probability distribution $P(\left\{n_{i}\right\})$ of having a certain
configuration with $n_{i}$ particles at site $i=1,\dots,L$. Explicitly, 
\begin{equation}
P(\left\{ {n_i}\right\}) =Z^{-1} \prod_{j=1}^ L f(n_j), 
\label{eq:fact}
\end{equation} 
where $f(n)$ depends on  the hopping rate. The factor $Z$ assures
normalization of the probability distribution, depends on the 
length of the lattice $L$ and the number of particles $N$ in the system.
For example, if $d_L(n) =p d(n)$ and   $d_R(n)=q d(n)$  with arbitrary 
$p>0$  and $q >0$, then 
\begin{equation}
f(n) = \prod_ {m=1}^{n}  \frac{1}{d(m)}.
\end{equation}
If the rate $d(n)$ goes to zero as $n\to\infty$, the system shows 
condensation, i.e., in the steady state there will be a finite number of
sites with a macroscopic occupation number~\cite{evan05}.

We will now analyze the case of non-vanishing $u$ and $v$,
where we focus on the case of one kind of particles. 
First, we will study a case that allows for an analytic solution
of the steady state probability distribution and shows that the 
the model can show condensation. Then we discuss the general
case by using a mean-field approximation.

\section{Factorized steady state}

Consider the case of $N$ particles of one kind that move 
on a periodic lattice of $L$ sites. The hopping rates are  
specified by 
\begin{eqnarray}
\label{eq:u}
u(m,n) &=& r w(m)w(n)  \\
\label{eq:v}
v(n) &=& rw(n)w(n-1)  \\
d_L(n) &=& q w(n) \\
\label{eq:dr}  
d_R(n) &=&  pw(n)
\end{eqnarray} 
where $w$ is an arbitrary  function of $n$ with $w(0)=0$ and 
$p$, $q$, and $r$ are positive parameters.  
As can easily be checked, these rates fall into the class introduced 
in the previous section. For this process, 
the steady state probability distribution factorizes and has the form 
(\ref{eq:fact}). Indeed, the steady state of the probability distribution 
$P$ is determined by
\begin{eqnarray}
\sum_{i=1}^{L}\left[2 u(n_{i-1}+1,n_{i+1}+1)P(\ldots,n_{i-1}+1,n_{i}
-2,n_{i+1}+1,\ldots)\right.\nonumber \\ 
-\{ u(n_{i-2},n_i) +u(n_{i},n_{i+2}) + 2v(n_i)\}  P(\ldots,n_{i-1},n_{i},n_{i+1},\ldots)\nonumber\\
+v(n_{i-1}+2) P(\ldots,n_{i-2}-1,n_{i-1}+2,n_{i}-1,\ldots) \nonumber\\
+v(n_{i+1}+2) P(\ldots,n_{i}-1,n_{i+1}+2,n_{i+2}-1  \ldots) \nonumber\\
+d_{R}(n_{i-1}+1)P(\ldots,n_{i-1}+1,n_{i}-1,\ldots)-
d_{L}(n_{i})P(\ldots,n_{i-1},n_{i},\ldots)&&\nonumber\\
\left. +d_{L}(n_{i+1}+1)P(\ldots,n_{i}-1,n_{i+1}+1,\ldots)-
d_{R}(n_{i})P(\ldots,n_{i},n_{i+1},\ldots)\right] &=&0 .\nonumber \\
\end{eqnarray} 
This condition is clearly fulfilled if each process occurs with the
same rate as the corresponding opposite process.
Inserting the ansatz (\ref{eq:fact}), this detailed balance 
yields $3L$ conditions
\begin{eqnarray}
u(n_{i-1}+1,n_{i+1}+1)f(n_{i-1}+1)f(n_{i}-2)f(n_{i+1}+1) &
= & v(n_{i})f(n_{i-1})f(n_{i})f(n_{i+1})\nonumber \\
d_{L}(n_{i+1}+1)f(n_{i+1}+1)f(n_{i}-1) & = & d_{L}(n_{i})f(n_{i-1})f(n_{i})
\nonumber \\
d_{R}(n_{i-1}+1)f(n_{i-1}+1)f(n_{i}-1) & = & d_{R}(n_{i})f(n_{i})f(n_{i+1})
\nonumber \\
\end{eqnarray}
with $i=1,\ldots,L$. For the special choices (\ref{eq:u})-(\ref{eq:dr}), 
they are solved by
\begin{equation}
w(n+1)\frac{f(n+1)}{f(n)} = const.
\end{equation}
Without loss of generality we choose the constant to be 1 and find
\begin{equation}
f(n) = \prod_ {m=1}^{n}  \frac{1}{w(n)}
\end{equation}
which formally is the same expression as for the ZRP.

The similarity of the steady state probability distribution with that
of the ZRP allows us to immediately carry over a number of results
to our system.
Of particular interest in the present context are the conditions for
which the particles condensate. Condensation means that in the 
``thermodynamic limit'', where
$N,L\to\infty$ with $ \rho= N/L=constant$,
a single site is occupied by macroscopic number of particles. 
Three cases can be distinguished as follows :
\begin{enumerate}
\item
$w(n) \to  \infty$ as $ n \to \infty$: No condensation,
the distribution of particles will be homogeneous in the limit $N\to\infty$.
\item
$w(n) \to 0$ as $ n \to \infty$: Condensation occurs 
for any density.   
\item
$w(n) \to\gamma>0$  as $ n \to \infty$ : 
Condensation depends on the first order correction to $w(n)$.
For example\footnote{In general, the first order correction to 
$\gamma$ is not $O(1/n)$. A detailed study of 
these non-trivial cases can be found in  
\cite{evan05}.}, if 
\begin{equation}
w(n) = \gamma\left(1 + b/n + O(1/n^2) \right)  
\end{equation}
then condensation occurs for  large enough densities 
iff $b> 2$.
\end{enumerate}   

Let us discuss case 3 further. First of all we note that
the leading term of $w$ implies that the process (a), see 
Fig.~\ref{fig:process}, occurs with a rate proportional to 
$n_{i-1}n_{i+1}$. This is reminiscent of the filament current 
(\ref{eq:jmimo}) which is proportional to the product of the filament
density at two different locations.
To discuss this case further, we will first consider 
\begin{equation}
w(n)= \gamma \left(1+ \frac{b}{n}\right)
\label{eq:wn}
\end{equation}  
where the higher order terms  are exactly $0$. Then 
\begin{equation}
 f(n) = \prod_{k=1}^n \frac{n\gamma^{-1}} {n+b} 
\end{equation}
which is asymptotically $f(n) \approx  \gamma^{-n} n^{-b}$. 

\begin{figure}
\centerline{ \includegraphics*[width=12 cm]{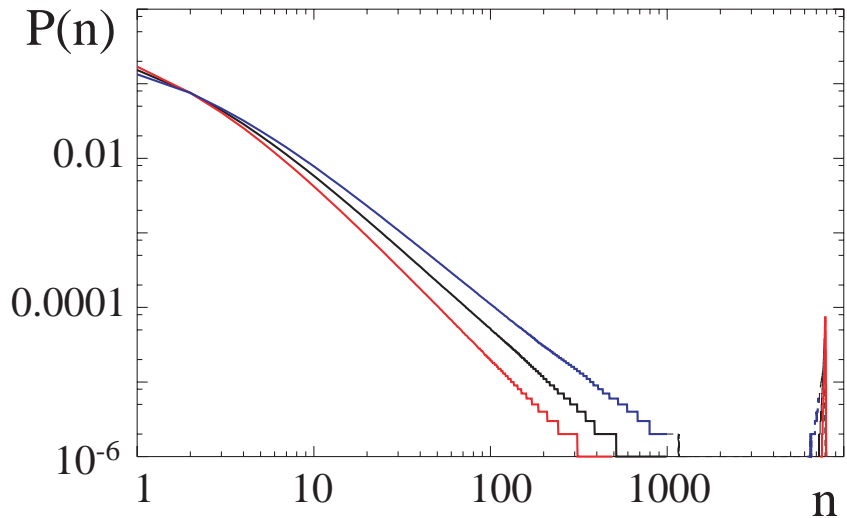}}
\caption{Probability $P(n)$ of finding in the steady state a site with $n$
particles obtained from stochastic simulations. The different curves 
correspond to different models within the general class where the
non-zero rates are specified by Eqs.~(\ref{eq:u})-(\ref{eq:dr}). 
Top: ZRP ($u=v=0$), Middle: Exactly solvable case (all rates 
non-zero), and  Bottom: $v=0$. In all cases, the probability distribution
shows first an algebraic decay and then a high probability for finding
a site containing a high fraction of the particles. The non-vanishing 
parameters are in 
all cases $r=0.24$, $p=0.6$, $q=0.4$ and $w(n)=1+ 2.5/n $. The 
length of the system was $L=2000$ and the number of particles were 
$N=6 L$.}
\label{fig:num}
\end{figure}

The distribution 
must respect the global conservation of particles, i.e., 
$\sum_i n_i = N$. It is helpful to consider the system in the grand canonical 
ensemble, where the density $\rho$ is controlled by the 
fugacity $z$. The grand partition function is 
\begin{equation}
F(z) = \sum_n \left( \frac{z}{\gamma}\right)^n f(n) 
\end{equation}  
which is well-defined  for $ 0\le z\le z_c$, where $z_c$ is the radius 
of convergence of $F(z)$.
For the rate (\ref{eq:wn}) one gets  
$z_c=\gamma$~\cite{evan05}. Knowing that the density of particles 
in the 
grand canonical ensemble is $ \rho(z) = \langle n\rangle  = 
z\frac{d \ln Z}{dz}$, the condensation can be  understood as follows:
if  $\rho$ diverges for $z\to z_{c}$ then  one can  obtain any arbitrary 
density by tuning $z$. However, if $\rho(z_c)<\infty$ then the 
maximum achievable macroscopic density is  just $\rho_c= \rho(z_c)$.  
Thus, in a system having density $\rho>\rho_c$, the extra 
particles,  in total $(\rho-\rho_c)L$, must form  the macroscopic 
condensate. 
For the rate (\ref{eq:wn}) it can be shown analytically~\cite{evan05} 
that $\rho_c= \frac{1}{b-2}$ is finite for $b>2$ . For any other form of 
$w(n)$ which asymptotically reduces to Eq.~(\ref{eq:wn}), for example
$\tilde w(n) = \gamma(1+ b /(n+c) )$,  $b_c$ is still $2$  whereas $\rho_c$ 
is in general different from $\frac{1}{b-2}$. When $b\ge 2$ 
and $\rho>\rho_c$, 
the probability of finding $n$ particles at a site decays
as $n^{-b}$ with average density $\rho_c$. The condensate occurs on 
top of this ``critical fluid''. See Fig.~\ref{fig:num} for an example. 
 
Thus, $b$ is the important parameter of the model which decides if the system 
can possibly phase seperate.  Physically, in the ZRP,  $b$ can be 
interpreted as the escape rate of particles from the condensate~\cite{evan05}. 
Note that the mean rate at which particles hop from the "critical fluid" is given by 
$\langle w\rangle = \gamma$ and thus particles escape from the condensate with 
a rate $w(n)-\langle w\rangle = \gamma b/n$.
   
Comparing these results to the minimal model of active filament
bundles we thus find a remarkable difference. While condensation
depends 
in both cases
on a critical density, 
in the hopping model 
the value of this critical density does not depend on the leading order 
term, but on the first order correction $b$.
In contrast, the critical value of the minimal model depends on
$\alpha$ which corresponds to the leading order term as will be shown
below.

\section{Absence of particle repulsion $v=0$ : Mean-field analysis}
\label{sec:mf}
\begin{figure}
 \centerline{\includegraphics*[width=12 cm]{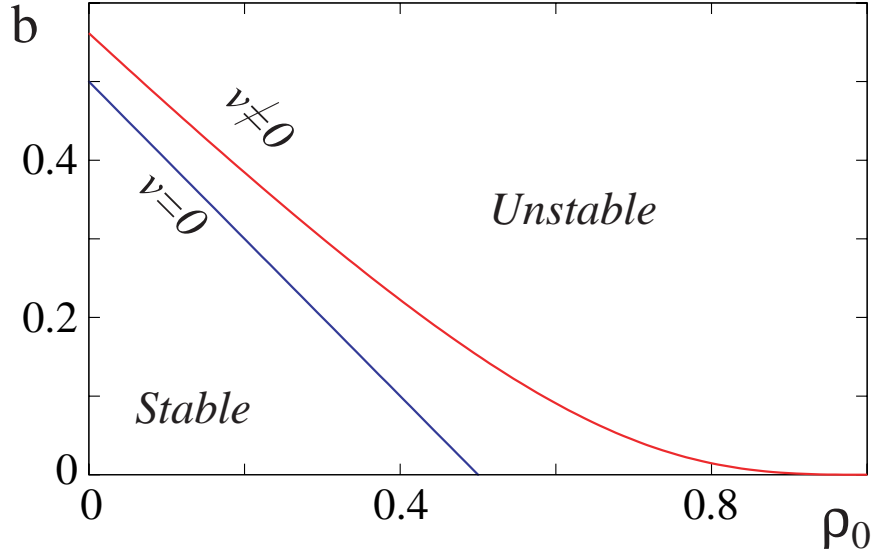}}
\caption{Mean-field analysis of the stochastic model with rates
specified in Eqs.~(\ref{eq:u})-(\ref{eq:dr}) with $w(n) = 1+ b/n$. 
The homogeneous state is stable for values of $b$ below the blue 
line if $v=0$ and below the red line if $v\neq0$. The parameter values 
are $p=q=r=1$, $\gamma=1$, and the
limit $L\to\infty$ is considered. } 
\label{fig:phase}
\end{figure}
In order to further investigate the last point we will now consider
the case $v=0$ when there is no repulsion of particles. In this
case an analytic solution is out of reach. Numerical solution reveals 
condensation 
for  $b>2$. As an example in  Fig.~\ref{fig:num} we show the 
steady state  distribution of particles for $b=2.5$. It shows a clear peak 
at high particle numbers similar to the corresponding distributions for
exactly solvable cases. For smaller particle numbers, the distributions
show a power law with an exponent $\xi=2.4 \pm 0.05$. The exponent 
is thus close to $b$, which is the value obtained in the exactly 
solvable case.

Some insight into the condensation behavior of the 
model can be gained by using a mean-field approximation.
To this end we first write the dynamic equations for the 
expectation values $\nu_i = \langle n_i \rangle$ for $i=1,\ldots,L$.
These equations depend on expectation values of products of
occupation numbers of different sites. The mean-field approximation
then consists of approximating these by products of the 
expectation values such that the dynamic equations are closed.
Following this procedure we find
\begin{eqnarray}
\frac{ d\nu_i}{dt} &=& d_{R}(\nu_{i-1})\nu_{i-1} - d_{R}(\nu_{i})\nu_{i} - 
d_{L}(\nu_{i})\nu_{i} + d_{L}(\nu_{i+1})\nu_{i+1} 
\nonumber  \\
&& +v(\nu_{i-1})\nu_{i-1}-2v(\nu_{i})\nu_{i}+v(\nu_{i+1})\nu_{i+1}
\nonumber\\
& &  -u(\nu_{i-2},\nu_{i})\nu_{i-2}\nu_{i} + 2u(\nu_{i-1},\nu_{i+1})\nu_{i-1}\nu_{i+1} -
u(\nu_{i},\nu_{i+2})\nu_{i}\nu_{i+2}
\label{eq:nu} 
\end{eqnarray}

First, consider again the rates (\ref{eq:u})-(\ref{eq:dr}) with $w$ given 
by Eq.~(\ref{eq:wn}), but now with $v=0$ and $p=q=r=1$.
It is easy to check  that $\nu_i= \rho_0= const.$ is  a  solution of 
(\ref{eq:nu}). Because of the global conservation of particles 
$L \rho_0= N$, thus $\rho_0$ must be the density of  the system. Stability 
of this homogeneous state $\nu_i=\rho_0$  can be 
checked by adding a small perturbation $\delta\nu_{i}$. Representing
the density by a Fourier-series we find for the dynamic equations 
in linear order in the perturbation
\begin{equation}
 \frac{ d\delta \nu_k}{dt} =  \epsilon_k  \delta\nu_k
\end{equation}
\noindent where
\begin{equation}
\epsilon_k = 2\gamma(\cos(2\pi k/L) -1) ( 1- 2\gamma (b+\rho_0) \cos(2\pi k/L)) 
\end{equation}
and  $\delta n_k$  are  the Fourier coefficients of the perturbation,
$k=1,\ldots,L$. Thus,
the pertubation $\delta \nu_k(t) = e^{\epsilon_k t} \delta \nu_k(0)$
decays if $\epsilon_k>0$ and grows in the opposite case.
Since  $\cos(2\pi k/L) -1)<0$ for all $k$,  modes 
can only become unstable if  $2(b+\rho_0 )\gamma>1$. In the limit 
$L\to\infty$ the instability occurs at $b_c= 1/2\gamma-\rho_0$. 
The minimal model discussed in section \ref{sec:minimal} is equivalent to  
the $b=0$ case  of Eq. (\ref{eq:wn}), where 
the critical value of $\gamma$ is inversely proportional to the 
density $\rho_0$. 

In the case when $v$ is chosen according to Eq.~(\ref{eq:v}) the 
critical value of $b$ is given by 
\begin{equation}
b_c = -(A^2 + 1/2) + \sqrt{ (A^2+1/2)^2 - 2A^3 + (1 - \gamma^{-1})A^{2} }, 
\end{equation}    
where $A= \rho_0 -1$.  Figure~\ref{fig:phase} describes the phase 
diagram  in $(b,\rho_0)$-plane for $\gamma=1$. Inclusion of the 
$v$-process shifts the critical line to larger values of $b$, that is, it
increases the stability region of homogeneous state.

\begin{figure}
 \centerline{ \includegraphics*[width=12 cm]{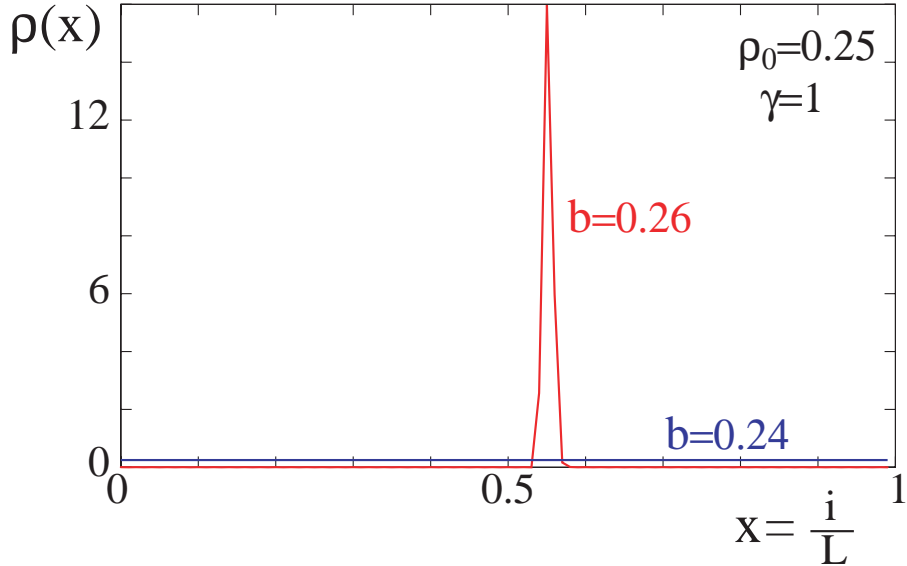}}
\caption{Steady state density profiles of the mean-field equation
(\ref{eq:nu}). Rates are given by (\ref{eq:u})-(\ref{eq:dr}) with 
$p=q=r=1$ and $w(n) = 1+ b/n$. Density profiles 
for a super- and a sub-critical value of $b$ are shown. Parameters
$\gamma=1$, $\rho_{0}=0.25$, such that $b_{c}=0.25$, and 
$b=0.24$ (blue line) and $b=0.26$ (red line).}
\label{fig:mf}
\end{figure}

A few points resulting from the mean-field analysis merit attention.
First, it is interesting that the condensation transition survives in the
mean field approximation,  
which is not the case  for ZRP, i.e.  when $v=0=u$. One can easily 
check that in this case $\epsilon_k=2\gamma(\cos(2\pi k/L) -1) \le 0$  
and thus $b_c=\infty$. Second, although the mean-field equations
of our model truly capture the transition, they do not reproduce the 
exact critical point $b_c=2$.  Also, the density profile obtained in the 
mean-field approximation is different from the exact results, see 
Fig.~\ref{fig:mf}. In the stochastic system, the condensate appears 
on the top of a critical fluid, whereas  the mean-field equations hide 
the fluctuations  coming from the fluid and all the matter is found in 
the condensate. It is also interesting to note  that the mean-field analysis 
would not change if $d$ is taken to be a constant $d=\gamma$ instead 
$d=\gamma(1+b/n)$, because first line of Eq. (\ref{eq:nu})  gives identical 
results for both cases. This interesting case $d=\gamma$ is further explored 
in next section.

\section{Connection to phenomenological descriptions of
active filament bundles}

Having discussed the driven diffusive system introduced in 
Section \ref{sec:model} for a specific choice of the rates, we now
want to return to the general case and to clarify the connection of
the stochastic models to the minimal model of active filament 
bundles which motivated our analysis. This will be done through 
the intermediate of phenomenological equations governing the 
dynamics of such bundles on large length scales.

Phenomenological theories of active gels - of which active bundles 
are a special example - are based on the observation that filament
currents are generated by gradients in the 
stress~\cite{krus03a,krus04,krus05,zumd05a}.
The system's stress can be expressed in terms of the state variables,
that is the dynamic fields describing the system.
If all filaments are aligned in a bundle and point all into
the same direction, the stress tensor reduces to a scalar $\sigma$
that depends on the filament density only. Furthermore, if any elastic
response of the system is neglected, then the relation between
the stress and the filament current $j$ can be written as
\begin{equation}
j=\eta\partial_{x}\sigma
\end{equation} 
where $\eta$ is an effective friction coefficient~\cite{krus03a}. In the 
phenomenological descriptions, the stress $\sigma$ is then 
systematically expanded in terms of the filament density and its
gradients with respect to some homogeneous reference state.

We can connect the hopping model to the phenomenological 
dynamic equation by considering a coarse grained version of the 
discrete mean-field equations of the hopping model. The 
coarse-graining procedure amounts to assuming that the occupation
numbers of the lattice sites vary only weakly with the site index
$i$. On large length scales the discrete index can then be replaced
by a continuous variable $x$ and the occupation numbers $\nu_{i}$
can be replaced by a density $c$ such that $c(x=ia)=\nu_{i}/a$. Here,
$a$ is the lattice spacing. Replacing $\nu_{i}$ in the discrete 
mean-field equations and then expanding terms like $c(x\pm a)$
into a Taylor-series $c(x\pm a)=c(x)\pm c'(x)a+c''(x)a^{2}\pm\ldots$,
which is truncated at some order of $a$,
we arrive at a partial differential equation for the time-evolution
of the density. 

We now apply this procedure to the mean-field equations of the
stochastic model ($i.e$, Eq. \ref{eq:nu}), where we focus for simplicity
on cases when only particles of one kind are present and when
$d_{L}=d_{R}=d$. Restricting the expansion to terms of at most
4th order in $a$ we find 
\begin{eqnarray}
\partial_{t}c & = &\partial_{x}^{2}\left[h(c)+\frac{1}{12}\partial_{x}^{2}
h(c)\right]\nonumber\\
&&- \partial_{x}^{2}\left[g(c,c)+\frac{7}{12}\partial_{x}^{2}g(c,c)-
2\partial_{1}\partial_{2}g(c,c)(\partial_{x}c)^{2}\right]
\end{eqnarray}
where without loss of generality we have set $a=1$ in the final
expression. In this equation, the hopping rates are contained in the
functions $h$ and $g$. Explicitly, $h(c) = d(c)c+v(c)c$ and 
$g(c,c)=u(c,c)c^{2}$. Furthermore, 
$\partial_{i}g(c,c)=\partial_{c_{i}}g(c_{1},c_{2})|_{c_{1}=c_{2}=c}$.
The dynamic equation for the filament density derived from the 
discrete mean-field equations of the hopping model are thus of the
same form as the corresponding equation for the filament density
in the phenomenological descriptions. From this we can deduce
the important result that the sum of the hopping rates  with appropriate 
sign 
multiplied by the filament density equals the 
tension $\sigma_{0}$ in the homogeneous state
\begin{equation}
\sigma_{0}= u(c_{0},c_{0})c_{0}^{2}-(d(c_{0})+v(c_{0}))c_{0}.
\end{equation}
This relation provides a simple link between the kinetic hopping 
rates, which are microscopic quantities, and the stress generated 
in the bundle, which is a macroscopic quantity. It should be noted,
however, that this relation is in general one way. While for every 
microscopic model the stress can be determined, the 
opposite is usually not true.

As an example consider the rates introduced in Section
\ref{sec:mf}, namely $d(c)=D$, $v(c)=0$,
and $g(c,c)=\gamma^{2}(c+b)^{2}$. In this case the dynamic
equation reads
\begin{equation}
\label{eq:mfcont}
\partial_{t}c = (D-2b\gamma^{2})\partial_{x}^{2}c+\frac{D-14b
\gamma^{2}}{12}\partial_{x}^{4}c-\gamma^{2}\partial_{x}^{2}\left[c^{2}-\frac{5}{6}
(\partial_{x}c)^{2}+\frac{7}{6}c\partial_{x}^{2}c\right]
\end{equation}
This equation can be compared to the expression obtained in the 
phenomenological approach developed in Ref.~\cite{krus03a}. This 
comparison shows that all terms appearing in the general expansion 
are generated by the stochastic model. It can therefore be used to
study the effects of fluctuations on the dynamics of a generic active 
filament bundle. Note, that in Ref.~[15] the term $\propto\partial_{x}^{4}c$
was neglected for reasons of simplicity.

A linear stability analysis of Eq.~(\ref{eq:mfcont}) yields a critical value $b_{c}$
of the parameter $b$ for which the homogeneous distribution loses
its stability. This corresponds to condensation in the stochastic model.
For the critical value we find $b_{c}=D/2\gamma^{2}-c_{0}$. This is
the same value as obtained in Section \ref{sec:mf} for $L\to\infty$ in the
mean-field limit of the discrete hopping model, where
$D=\gamma$. The corresponding tension  created in  the homogenous state
at the critical point can be expressed as $\sigma_{0}=\gamma^{2}(c_{0}+b)(b+c_{0}-2)$. 
An instability of the homogenous state leads to a state where filaments condense,
i.e., accumulate at one point and can therefore be interpreted as the point where
the bundle ruptures. The value of the tension at the critical point is thus the maximal 
tension that can be generated by a filament bundle before it tears apart. 

\section{Discussions and conclusions}

In summary, we have introduced a new class of hopping models
on one-dimensional lattices. Motivated by the dynamics of filaments
connected by molecular motors, this class is defined by
processes in which two particles hop simultaneously; either
from adjacent sites to a site in between or away from one site to
both adjacent sites. This is, in particular, different from the
ZRP where such processes occur in an interval $\Delta t$ with
a probability $\propto(\Delta t)^{2}$ and are therefore neglected
in comparison to processes that occur with a probability 
$\propto\Delta t$. 
As a consequence of the two-particle processes the mean-field
approximation shows like the stochastic system a condensation
transition. This does not hold for the mean-field analysis of the ZRP.
Despite these important differences, for rates that 
have a factorized steady state probability distribution, the steady
state distributions of the ZRP and of the class of systems introduced
here are identical. This similarity allowed us to transfer
results from the ZRP on the conditions of condensation to the 
new model class. In particular, we found that there is a critical 
density for condensation, if the rates approach for large occupation
numbers a constant different from zero.

Coarse-graining of the mean-field equations allowed us to connect the
kinetic hopping rates in the stochastic model to the stress generated
in the bundle. The microscopic rates are thus linked to macroscopic
physical quantities. Apart from giving the rates a physical meaning,
this connection
could be used to obtain the strength of the noise term in Langevin
equations for the time-evolution of the filament density. As active
filament bundles are inherently out of thermodynamic equilibrium
there is in general no fluctuation-dissipation theorem that would 
allow to determine this strength. The coarse-graining also points to 
limitations of continuum descriptions of active
filament bundles. Indeed we find that the criterion for condensation
in the exactly solvable case involves only the first order correction
of the rates, while the critical value in the mean-field equations 
depends also on the zeroth order parameter. 

Taken these points together, the discrete hopping models introduced
here are seen to be a useful tool for analyzing the effects of 
fluctuations on dynamics of
active filament bundles - in spite of their simplicity. In the future it
will be in particular interesting to study the case of two different
particles present on the lattice. In the corresponding microscopic
and phenomenological descriptions, traveling waves were found in
this case.

\end{document}